\documentclass[aps,prb,twocolumn,preprintnumbers,amsfonts,a4paper]{revtex4-1}

\usepackage{eurosym}
\usepackage{graphicx}
\usepackage{float}
\usepackage{placeins}

\usepackage[utf8]{inputenc}

\usepackage[sort&compress]{natbib}

\usepackage{bm}          
\usepackage{amsmath}
\usepackage{latexsym}
\usepackage{color}

\graphicspath{{./}}

\newcommand{\HH}{{\cal H}}
\newcommand{\HHF}{\HH_{\mathrm{HF}}}
\newcommand{\HKS}{\HH_{\mathrm{KS}}}
\newcommand{\HiMF}{\HH_{\mathrm{iMF}}}

\newcommand{\vF}{{v_{\mathrm F}}}

\newcommand{\ket}[1]{{|{#1}\rangle}}
\newcommand{\bra}[1]{{\langle{#1}|}}

\newcommand{\Eaverage}[1]{\ensuremath{\langle{#1}\rangle}}

\newcommand{\Eqref}[1]{Eq.~\eqref{#1}}
\newcommand{\Figref}[1]{Fig.~\ref{#1}}

\begin{document}

\title{Inverse Mean Field Theory}

\author{Peter Schmitteckert}
\affiliation{Institute for Theoretical Physics and Astrophysics,
Julius-Maximilian University of W\"urzburg,
Am Hubland, 97074 W\"urzburg,
Germany}
\date{\today}

\begin{abstract}
In this work we discuss the extraction of mean field single particle Hamiltonian from 
a many body wave function of a fermionic system. 
It allows us to discuss the result
of a many particle wave function in terms of a non-interacting description.
\end{abstract}

\maketitle

\section{Introduction}
%
Many particle wave function based methods, like the density matrix renormalization group (DMRG) approaches%
\cite{White:PRB93,White:PRL92,Proceedings98,Schollwoeck:RMP2005,Noack_Manmana:AIPCP2005,Hallberg:AP2006}
are ideally suited to tackle strongly correlated quantum  systems. However, due to their many particle wave functions
formulation it is hard to discuss the results beyond the measurement of observables. In contrast, mean field theories,
i.e. single particle based descriptions, can be quite problematic in describing strongly correlated systems.
Yet, they are much easier to discuss, as one can i the single particle levels and their occupations.
Ideally one would like to have a single particle based description describing the essence of the underlying physics.
We therefore pose the question whether it is possible to obtain a suitable single particle description out of the many
particle solution. 

The situation is comparable to scattering theory. There one seeks for the scattering matrix corresponding to a given
Hamiltonian. Then there are inverse scattering methods, most notably the quantum inverse scattering method \cite{Korepin:QISM1993}
(QISM).
There one  constructs an Hamiltonian $\HH$ out of a scattering matrix ${\cal S}$
Within the so-called algebraic Bethe ansatz one starts with a scattering matrix
fulfilling the Yang Baxter equation, essentially requiring the many particle scattering process being independent of the order
of the scattering processes. One can then construct a corresponding many particle Hamiltonian, that is  solved by the initial
scattering matrix. Here we want to invert the traditional way of mean field theory where one starts with a
single particle approximation to an Hamiltonian which then leads to a single particle wave function as solution.
Instead we start from the many particle wave function and extract a single particle Hamiltonian from the wave function.
In that sense we call this approach inverse mean field theory, since we are inverting the direction from the wave function 
to the Hamiltonian.

We start by demonstrating the idea with an inverse Hartree Fock (iHF) approach. We then summarize inverse density functional
theory (iDFT), which we then extend in order to include off-diagonal matrix elements.

\section{Inverse Hartree Fock}
%
The Hartree-Fock (HF) approximation represents an important technique in studying interacting systems.
For simplicity we discuss the HF method with spinless fermions, or alternatively, the spin index gets
absorbed into the site index. 
Starting with a general Hamiltonian featuring density density interaction
\begin{align}
	\HH =& \sum_{x,y} h_{x,y} \hat{c}^+_x \hat{c}_y  \;+\; \sum_{x,y}  U_{x,y} \hat{n}_x \hat{n}_y  \,,
\end{align}
with $\hat{c}^+_x$ ($\hat{c}^{}_x$) the standard canonical fermionic creation (anihilation) operators at site $x$.
Restricting the space of solutions to single particle Slater determinants 
we obtain a single particle approximation $\HHF$ 
\begin{align}
    \HHF =& \sum_{x,y} h_{x,y} \hat{c}^+_x \hat{c}_y   \label{eq:HF_SP} \\
						& + \sum_{x,y}  V_{x,y} \left(\, \Eaverage{ \hat{n}_x}  \hat{n}_y +  \Eaverage{ \hat{n}_y}   \hat{n}_x \,\right)  \label{eq:HF_H}  \\
						& - \sum_{x,y}  V_{x,y} \left(\, \Eaverage{ \hat{c}^+_x \hat{c}^{}_y }   \hat{c}^+_y \hat{c}^{}_x  + \Eaverage{ \hat{c}^+_y \hat{c}^{}_x }   \hat{c}^+_x \hat{c}^{}_y \,\right)  \label{eq:HF_F} \\
						& - \sum_{x,y}  V_{x,y} \left(\, \Eaverage{ \hat{n}^{}_x}  \Eaverage{\hat{n}^{}_y  } -  \Eaverage{ \hat{c}^+_x \hat{c}^{}_y }  \Eaverage{ \hat{c}^+_y \hat{c}^{}_x} \,\right)  \label{eq:HF_DC} 
\end{align}
consisting of the Hartree contribution $\HH_{\mathrm{H}}$ \eqref{eq:HF_H},
the Fock or exchange contribution  $\HH_{\mathrm{F}}$ \eqref{eq:HF_F}, and a compensating term, \Eqref{eq:HF_DC}, accounting for the double counting of interaction terms.
In a self-consisting HF calculation of typically starts with diagonalizing the non-interacting Hamiltonian. One then calculates the necessary expectation values
appearing in Eqs.~(\ref{eq:HF_H},\ref{eq:HF_F},\ref{eq:HF_DC}) and determines a new single particle Hamiltonian $\HHF$.
One then repeats calculating expectation values and constructing a new Hamiltonian $\HHF$ until the loop converges.
In order to avoid oscillating behavior it is advisable to take a weighted average of   old and new parameters. In the simulations reported here
we weight the new expectation values with $30\%$. 

\subsection{Friedel oscillations}
\label{sec:FO}

\begin{figure}[t]
\includegraphics[width=0.9\columnwidth,type=pdf,ext=.pdf,read=.pdf]{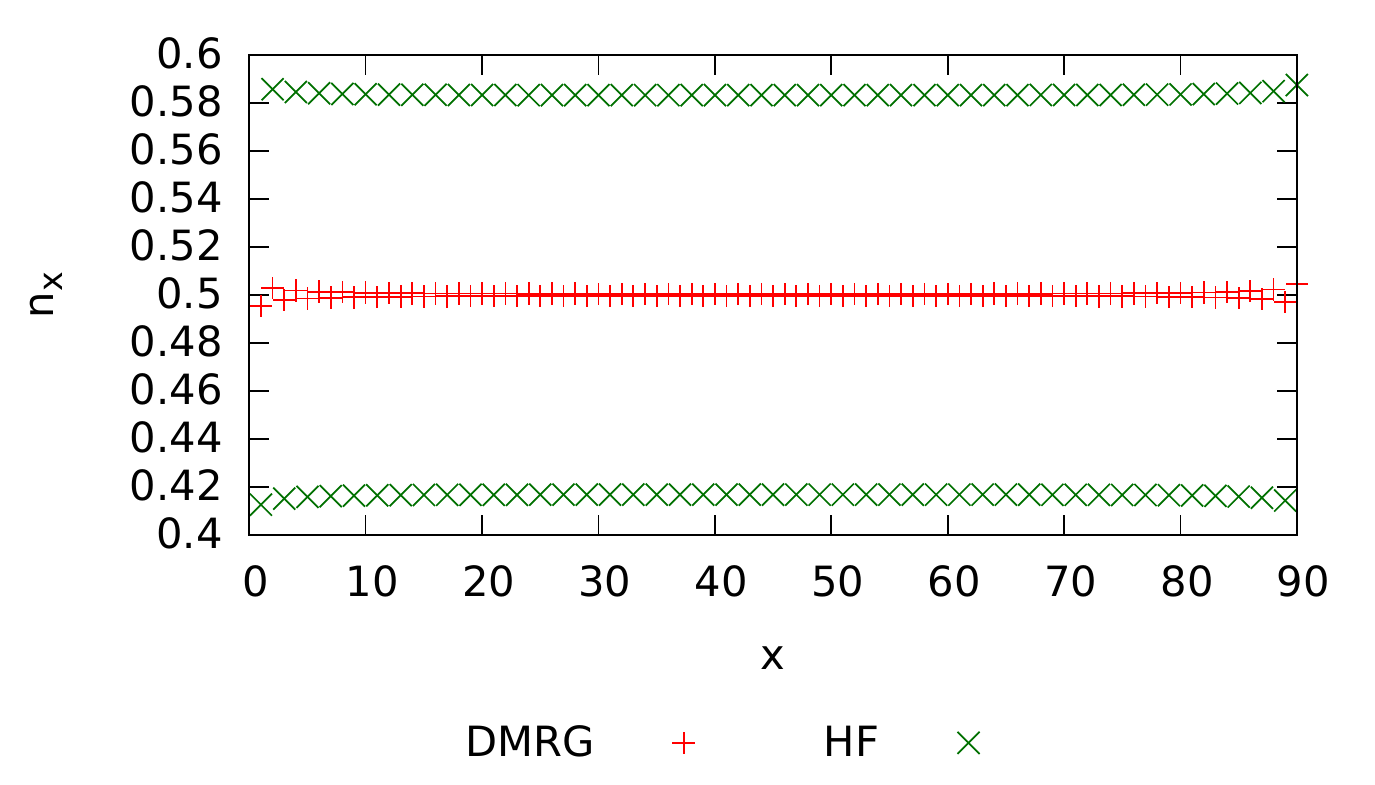}
\caption{Local densities of a one--dimensional fermionic $M=90$ site system with a nearest hopping of $t=1$,
nearest neighbor interaction $U=1.0$, periodic boundary conditions and a local potential $\epsilon=\pm 0.01$
on the first and last site of the system. The system is solved via DMRG (plusses) and SCHF (crosses).}
\label{fig:HF}
\end{figure}
\begin{figure}[!t]
\centerline{\includegraphics[width=0.9\columnwidth,type=pdf,ext=.pdf,read=.pdf]{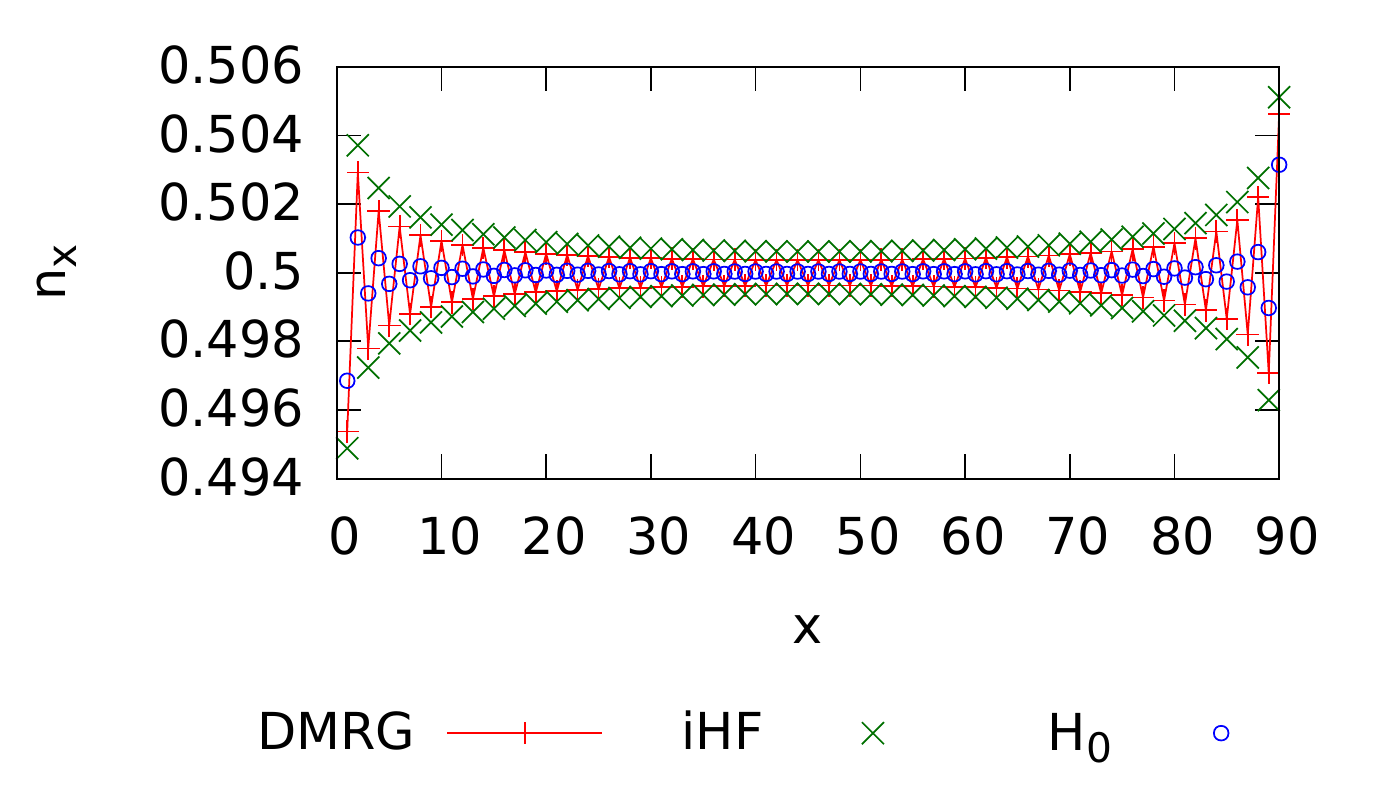}}
\caption{inverse Hartree Fock --- Here we display the local densities for the same system as in \Figref{fig:HF}.
 The local densities are obtained from DMRG (pusses and line), the non--interacting system corresponds to $U=0$ (circles),
  and iHF (crosses) is obtained from  the HF Hamiltonian where the  expectation values defining the HF Hamiltonian are extracted from the DMRG ground state wave function.}
\label{fig:iHF}
\end{figure}

In \Figref{fig:HF} we compare the local density obtained from a DMRG simulation of a one--dimensional fermionic system
\begin{align}
  \HH =& -t \sum_{x=1}^M \hat{c}^+_{x-1} \hat{c}^{}_x  \,+\,\hat{c}^+_{x} \hat{c}^{}_{x-1} \;+\ U \sum_{x=1}^M \hat{n}^{}_{x-1} \hat{n}^{}_x  \nonumber \\
       & + \epsilon \left( \hat{n}^{}_1 - \hat{n}^{}_M \right)
\end{align}
with a nearest neighbor hopping on $t=1.0$, a nearest neighbor density density interaction with strength $U=1.0$,
zero local potentials except on the first and last site where we employ a local potential of $\pm \epsilon$, $\epsilon=0.01$,
and periodic boundary conditions (PBC). Within the DMRG using our own code we kept 1500 states per block, 11 finite lattice sweeps,
leading to discarded entropies, i.e.\ a measure for the information thrown away at each DMG step, of below $10^{-10}$.
Within this work we can view these DMRG results as being the exact values.
The HF self consistency loop was iterated until the changes where below $10^{-10}$.
As one can clearly see, the induced Friedel oscillations are far too large within HF leading to an ordered state, while the 
true ground state is a Luttinger liquid. This fits to common believe that HF does not work for 1D interacting fermions.
In addition, Friedel oscillations of an interacting one--dimensional wire show a clear signature of Luttinger liquid behaviour,
as the decay of the envelope of the oscillations is a pow law with an interaction dependent exponent \cite{Egger:PRL1995,PS_Eckern:PRB1996,Lesage_Saleur:JPA1997}. 
This is in strong contrast
to non-interacting systems, where the decay is proportional to  $1/r^d$ with $r$ the distance from the impurity and $d$ the dimensionality
of the system \cite{Friedel:NCS1958,Zawadowski:PRB1985}.
In contrast we show in \Figref{fig:iHF} the same system as in  \Figref{fig:iHF} with the exception that the self--consistent HF (SCHF)
is replaced by an inverse HF (iHF), where instead of calculating the expectation values needed for the HF calculation 
self--consistently from the HF itself, we measured the corresponding expectation values in the DMRG solution and used them,
without any further HF loops. 
We immediately see that now the Friedel oscillations are much better reproduced displaying
only a slight over--shooting. We would like to point out that this result is not obvious, as these iHF densities are obtained from
a single slater determinant of the non-interacting iHF system, while the true system corresponds to  Luttinger liquid with
non-trivial correlations.

%
In addition these results hint at the reason of the typical failure of the SCHF.
If we would start a self--consistency loop from our iHF we will actually increase our HF parameter leading to stronger and stronger deviations.
Therefore, it's the self-consistency loop that leads to the breakdown of the SCHF shown in \Figref{fig:HF}.

\section{Inverse DFT}
Density functional theory (DFT) is one of the most used numerical technical techniques to study solid state systems.
It is based on the Hohenberg-Kohn theorem \cite{Hohenberg_Kohn:PRB1964} stating that for every observable $\cal O$ of an electronic system,
there is a density functional providing the expectation value $\Eaverage{\cal O}$ within the ground state provided it is
evaluated with the ground state density. That is, the language changes from the wave functions to density functionals
and is typically studied within a Kohn--Sham auxiliary system \cite{Kohn_Sham:PR1965}.
There one replaces  the Hamiltonian $\HH$ of an interacting fermionic system
\begin{equation}
\HH \;=\; \underbrace{\sum_{\ell \ne m} t_{\ell,m} \,\hat{c}^+_\ell \hat{c}^{}_m \,+\,\text{h.c.}}_{\text{kinetic part: } \cal K} 
	\;+\; \underbrace{\HH_{\mathrm{int}}}_{\text{interaction}}
	\;+\; \underbrace{\sum_\ell V^{\mathrm{ext.}}_\ell \hat{n}_\ell}_{\substack{\text{ext. potential}}}
\end{equation}
with $t_{\ell,m}$ the hopping parameter providing the kinetic part $\cal K$, the interacting part $\HH_{\mathrm{int}}$  including all non-single particle terms,
and $V^{\mathrm{ext.}}_\ell$ local potentials,
with an non interacting Hamiltonian $\HKS$
\begin{equation}
\HKS \;=\; {\cal K}  \,+\,\sum_\ell v^{\mathrm{HXC}}_\ell \hat{n}_\ell \,+\, \sum_\ell V^{\mathrm{ext.}}_\ell \hat{n}_\ell
\end{equation}
with the same kinetic part $\cal K$ and external potentials $V^{\mathrm{ext.}}_\ell$. However, the interacting part $\HH_{\mathrm{int}}$
is replaced by so--called Hartree--exchange--correlation (HXC) potentials $v^{\mathrm{HXC}}_\ell$. The HK theorem now guarantees that there is
a unique HXC potential such that the ground state density $n_0$ of the non--interacting  Hamiltonian $\HKS$ is the same as the the ground state
density of the fully interacting system.

Within inverse density functional theory (iDFT), or more accurately inverse site occupation function theory (iSOFT),
being introduced by \textcite{SchoenhammerGunnarsson:PRL1986} and extended to inhomogeneous systems in \cite{Schmitteckert_Evers:PRL2008},
one now starts from the many particle ground state wave function and determines the local $v^{\mathrm{HXC}}_\ell$ potentials,
such, that the ground state density of the auxiliary system is the same as in the interacting system. In this way one can construct 
the HXC potentials corresponding to the initial Hamiltonian. By construction we would now arrive at a single particle Hamiltonian
reproducing the Friedel oscillations of Figs.~(\ref{fig:HF},\ref{fig:iHF}). While this appears to be clearly superior to the iHF 
approach one has to take in account that the lattice formulation of DFT has one major draw back. The kinetic part is not expressed
as the gradient of the density, but by independent hopping parameter. And this can lead to major difficulties.
For this reason the existence of a time dependent version of the DFT is not guaranteed in a lattice \cite{Tokatly:PRB2011,PS_Dzierzawa:PCCP2013}
which is in contrast to the Runge--Gross theorem for continuous systems \cite{Runge_Gross:PRL1984}.

\section{Inverse Mean Field Theory}

We would therefore like to extend the DFT on the lattice to a reduced Density Matrix Functional (rDMFT) \cite{Gilbert:PRB1975,Coleman:RMP1963}.
There one extends the density functional to the complete single particle density matrix $K$ (SPRDM)
as the fundamental variable
\begin{align}
	K_{x,y} &= \bra{\Psi}\, \hat{c}^+_x \hat{c}_y \,\ket{\Psi} \,.
\end{align}
However, there is a fundamental problem. The eigenvalues of the SPRDM are not restricted to zero and one. In results the single particle density matrix
is not idem--potent and one can not resort to a Kohn--Sham auxiliary system for RDMFT \cite{Kohn_Sham:PR1965,Coleman:RMP1963,Gilbert:PRB1975,Levy:PNAS1979,Mueller:PLA1984,Sharma:PRB2008}in a similar manner as 
in DFT. Nevertheless we show in this work that one can extract single particle models in the spirit of the iHF that comes close to the desired goal.

To this end we study the single particle Hamiltonian 
\begin{align}
	\HiMF  &= \hat{c}^+_x \,H_{x,y} \, \hat{c}_y  \\
	H_{x,y} &= \bra{\Psi} \hat{c}^{}_x \left( \HH -E \right) \hat{c}^{+}_y \ket{\Psi} 
	 \,-\, \bra{\Psi} \hat{c}^+_x \left(\HH-E\right)\hat{c}_y \ket{\Psi} \,, \label{eq:DOOFT_H}
\end{align}
where $\ket{\Psi}$ denotes an eigenstate, typically the ground state,  of our interacting Hamiltonian $\HH$
with $E$ the corresponding energy eigenvalue.

Suppose our Hamiltonian of interest is actually non--interacting,
\begin{align}
 \HH  &= \sum_{x,y} \,H^0_{x,y}\, \hat{c}^+_x \hat{c}_y  \;=\; \sum_\ell \varepsilon_\ell \hat{n}_\ell \,. \label{eq:H0} \\
  E_0 &= \sum_\ell \varepsilon_\ell \,n_\ell \,,
\end{align}
with $n_\ell = \Eaverage{\hat{n}_\ell}$.
Evaluating \Eqref{eq:DOOFT_H} in the basis of eigenstates of \Eqref{eq:H0} and taking care that $n_\ell = \bra{\Psi} \hat{n}_\ell \ket{\Psi}$
is evaluated with respect to the reference state $\ket{\Psi}$  we obtain
\begin{align}
H_{p,q} &=  \bra{\Psi}\, \hat{c}^{}_p \,\left( \sum_\ell \varepsilon_\ell \left(\hat{n}_\ell - n_\ell\right) \right) \, \hat{c}^{+}_q \,\ket{\Psi}  \nonumber\\
        &-  \bra{\Psi}\, \hat{c}^+_p  \,\left( \sum_\ell \varepsilon_\ell \left(\hat{n}_\ell - n_\ell\right) \right) \, \hat{c}^{}_q \,\ket{\Psi}\\
        &=  \bra{\Psi}\, \hat{c}^{}_p \,\left(  \varepsilon_q \left(\hat{n}_q - n_q\right) \right) \, \hat{c}^{+}_q \,\ket{\Psi}  \nonumber\\
        &-  \bra{\Psi}\, \hat{c}^+_p  \,\left(  \varepsilon_q \left(\hat{n}_q - n_q\right) \right) \, \hat{c}^{}_q \,\ket{\Psi}\\
        &= \delta_{p,q} \epsilon_p \bra{\Psi}\, \left( 1- n_p\right)\hat{c}^{}_p \hat{c}^+_p \,+\, n_p\hat{c}^{+}_p \hat{c}^{}_p\, \ket{\Psi}\\ 
        &= \delta_{p,q} \epsilon_p \bra{\Psi}\, \left( 1- 2n_p + 2 n^2_p\right) \, \ket{\Psi}\\ 
        &=  {\delta_{p,q}\, \varepsilon_p } \,.
\end{align}
Where in the last step we assumed the SPDM $K$ to be idempotent, i.e. the occupation numbers to be restricted to zero and one.
In result, the construction \Eqref{eq:DOOFT_H} applied to any eigenstate $\ket{\Psi}$ of the non-interacting Hamitlonian $\HH$ 
will reveal the original Hamiltonian.
We take this observation as a motivation to define our iMF Hamiltonian $\HiMF$  via \Eqref{eq:DOOFT_H}. 

However, in contrast to non-interacting systems, specifically in the case that the reference state $\ket{\Psi}$ is not given by a single slater determinant,
the SPRDM $K$ and the iHF Hamiltonian Matrix $H$ will in general not commute.
  
While it would be most accurate to setup a single particle description based on the extracted $K$ and $H$,
we face the problem that it is inconvenient to work we a single particle theory where $K$ and $H$ do not commute.
and since the purpose of this work is to introduce a description where our usual concepts for non--interacting systems can be
applied we introduce the following approaches.

\begin{itemize}
\item {\bf i}nverse {\bf N}atural {\bf O}rbit {\bf O}ccupation {\bf F}unctions.
\begin{itemize}
		\item Diagonalize $K_{\ell,m} = f_\ell \,\delta_{\ell,m}$.
		\item Diagonalize $H$ in degenerate subspaces of $K$.\\
		  $\Rightarrow\, \HH^{\mathrm N}=\sum_\ell \varepsilon^{\mathrm N}_\ell \hat{n}_\ell$.
\end{itemize}
\item {\bf i}nverse {\bf D}ynamic {\bf O}rbit {\bf O}ccupation {\bf F}unctions.
\begin{itemize}
		\item Diagonalize $H_{\ell,m} = \varepsilon^{\mathrm D}_\ell \,\delta^{}_{\ell,m}$.
		\item Diagonalize $K$ in degenerate subspaces of $H$.\\
		  $\Rightarrow\, \HH^{\mathrm D}=\sum_\ell \varepsilon^{\mathrm D}_\ell \hat{n}^{}_\ell$.
\end{itemize}
\end{itemize}

Within the first approach, iNOOF, we diagonalize the SPDM $K$, transform $H$ in the same basis, the so-called natural orbitals,
and take only the diagonal term as matrix elements of out new Hamiltonian. In case $K$ has degenerate eigenvalues, we diagonalize
$H$ within that degenerate space. In result we have constructed a single particle Hamiltonian, that will lead to the same natural orbitals
as the interacting system.

In the second approach, iDOOF, we take the opposite approach. We first diagonalize $H$, and then diagonalize $K$ within the degenerate
eigenspaces of $H$ and take the resulting eigenstates as basis. Now we constructed a single particle Hamiltonian, that share the same single
particle excitations as the interacting one. In the following we will concentrate on this second approach and refer to it  as iMF.

In both approaches we can take the distribution function, i.e.\ the occupation of the single particle eigenstates, as extracted
from $\ket{\psi}$ or we can simplify our approach by approximating the occupation numbers by a $\Theta$ function.
While the definition of the inverse single particle Hamiltonian is straightforward and non--ambiguous it is not obvious that
the concept is actually useful. We therefore provide examples of the iMF the remaining sections showing that the approach can indeed
provide insight beyond standard DFT on the lattice. We would also like to point out that by tabulating the obtained mean field Hamiltonians
obtained by scanning the set of possible local potentials and non-local hopping terms, on can in principle construct a reduced density matrix function(al)
for the lattice models.

\subsection{Friedel oscillations}
\begin{figure}
 {\includegraphics[width=0.9\columnwidth,type=pdf,ext=.pdf,read=.pdf]{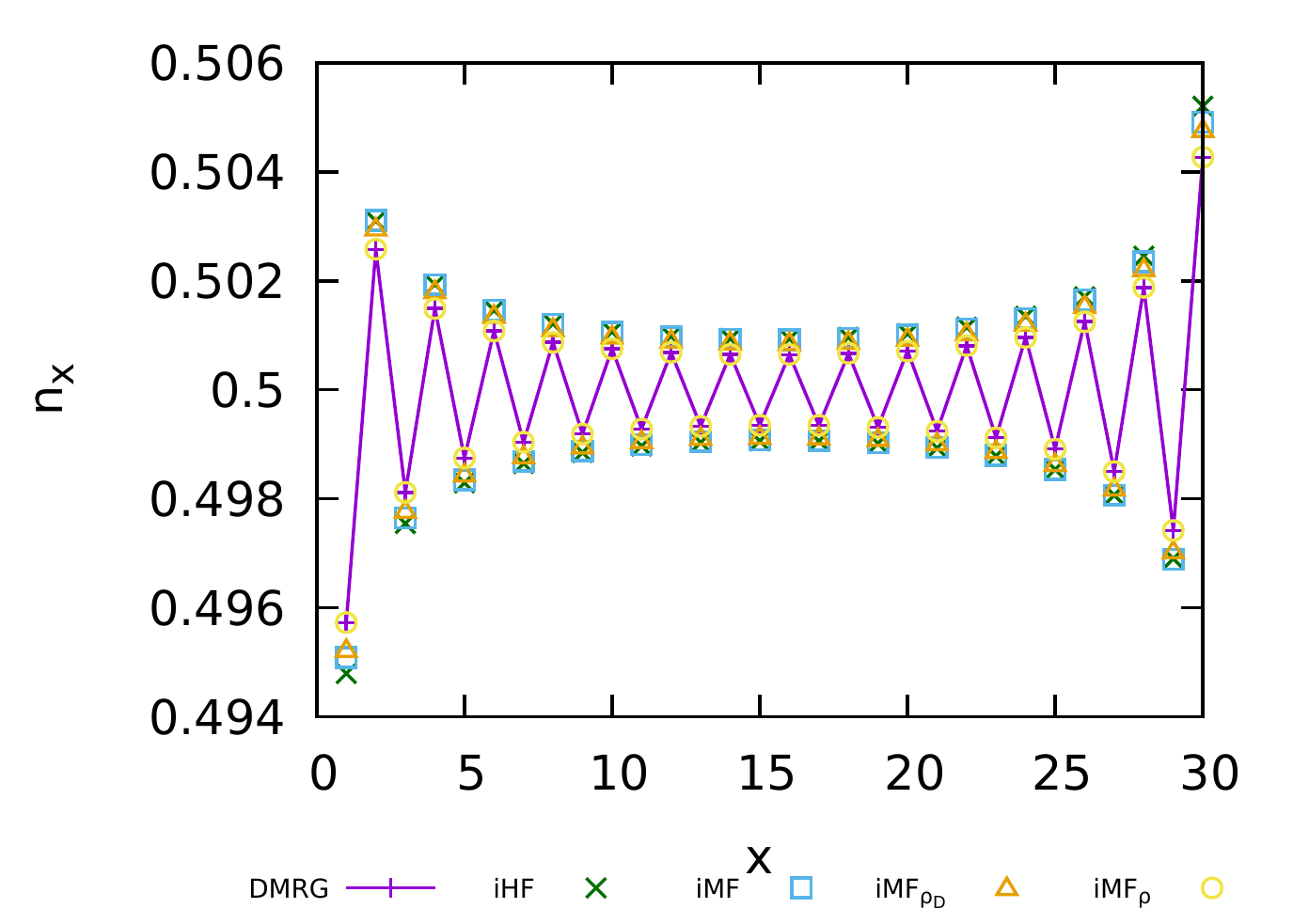}}
\caption{inverse mean field: Friedel oscillations --- Here we display the local densities for the same system as in \Figref{fig:HF}.
 The local densities are obtained from DMRG (plusses and line), the inverse HF results (crosses) correspond to the results shown in  \Figref{fig:iHF}.
 The iMF (squares) results is obtained from an iDOOF approach, where the occupation function is approximated by a $\Theta$ function.
 $\text{iMF}_{\rho_{\mathrm D}}$ (triangles ) is obtained from an iDOOF approach using the diagonal part of $K$ as extracted from the ground state
 and the $\text{iMF}_{\rho}$ (circles) data are obtained by resorting to the full SPDM $K$. }
\label{fig:iMF_Friedel}
\end{figure}

We start by applying the iMF to the problem of Friedel oscillations as introduced in iHF section.
In \Figref{fig:iMF_Friedel} we display the the Friedel oscillation for the same system as in in \Figref{fig:HF}.
The results from DMRG and iHF are the same as in \Figref{fig:HF} and \Figref{fig:iHF}.
With respect to the iMF we applied several versions of the iDOOF. First we extracted a single particle Hamiltonian as described in 
the iDOOF scheme and replaced the occupation function by a $\Theta$ function, $n_\ell=\Theta\left(-\varepsilon_\ell\right)$.
We find that the obtained densities are similar to the iHF approach except at the impurities, where the iMF provides slightly better results.
By taking the occupation function as obtained from the diagonal part of $K$ in the basis where $H$ in diagonal, we obtain the results
labeled with $\text{iMF}_{\rho_{\mathrm D}}$. Non surprisingly, these density values are systematically better compared to iHF.
Finally, we can also build a single particle approach taking $K$ and $H$ fully into account without restricting to diagonal parts.
Of course, we now recover the DMRG results by construction.

\subsection{Band structure}

In the previous section we have shown that iMF can provide accurate results for the Friedel oscillations being better than iHF.
However, iDFT would by construction provide the correct local densities.
In this section we show that the advantage of our iMF compared to iDFT is that it takes the kinetic part into account.

To this end we study the same interacting chain as in the Friedel oscillations problem, except that we do not apply local scattering potentials, i.e.\ $\epsilon_x=0$,
\begin{align}
\HH  =& -t \sum_x \hat{c}^+_{x-1} \hat{c}^{}_x  \,+\,\text{h.c.} \nonumber\\
      &+\; U \sum_x \left( \hat{n}_{x-1}-\frac{1}{2}\right) \left( \hat{n}^{}_x -\frac{1}{2}\right) \,. \label{eq:H_LL}
\end{align}
\begin{figure}[t]
{\includegraphics[width=0.9\columnwidth,type=pdf,ext=.pdf,read=.pdf]{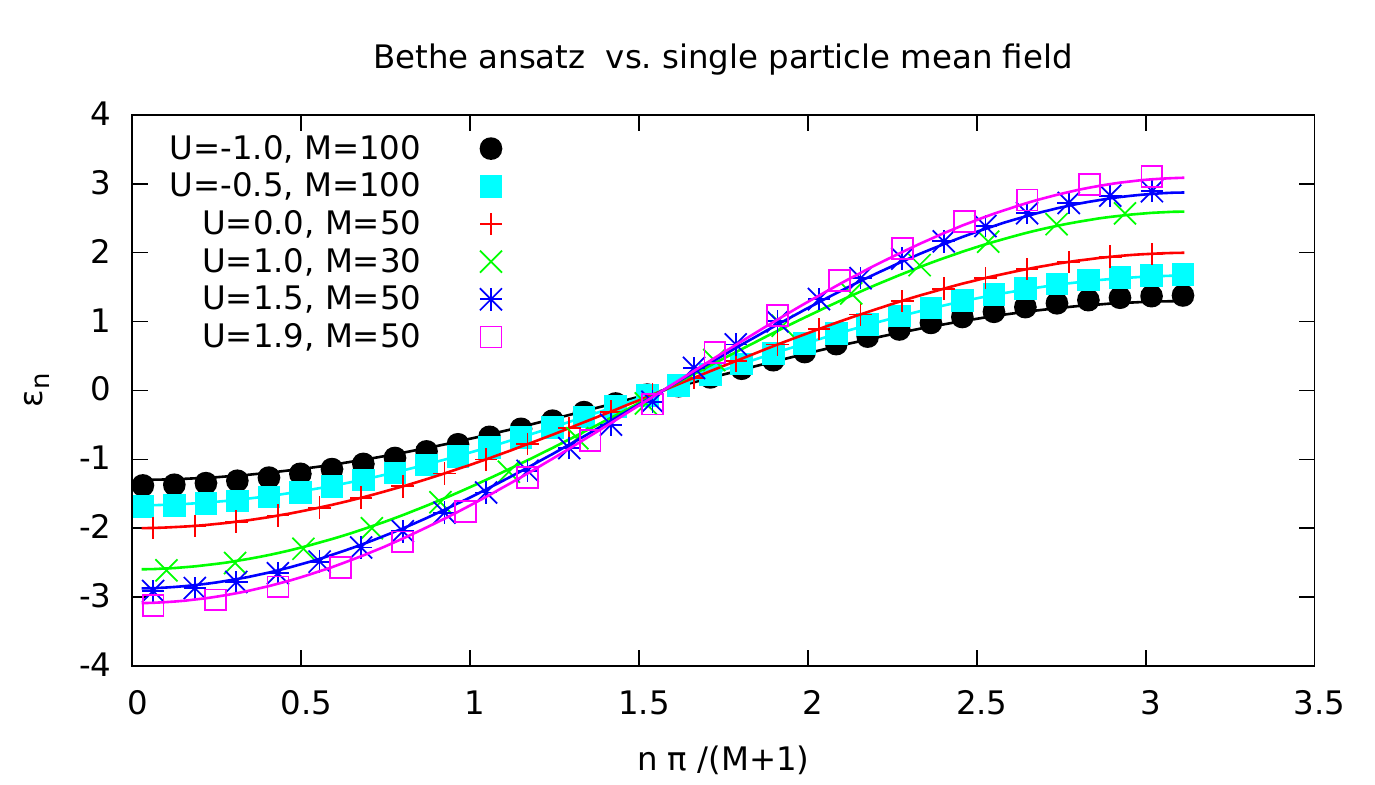}}
\caption{Band structure of the iMF system obtained from the ground state of one-dimensional Fermi system with nearest neightbor hopping of $t=1$,
 nearest neighbor interaction $U$ and hard wall boundary conditions. We plot the eigenenergies vs.\ the momentum values suitable for hard wall boundary conditions, 
 see Ref.~\cite{Schmitteckert:JPCS2010}.}
\label{fig:LL_vF}
\end{figure}

In \Figref{fig:LL_vF} we display die energy dispersion for obtained for several nearest neighbor interaction values. We applied hard wall boundary condition (HWBC)
as it double the resolution in momentum space, provide be evaluate the the energy dispersion at momentum values $k_\ell = \frac{\ell\pi}{ M+1}$, $\ell=1, 2, \cdots, M$, 
with $M$ the number of lattice sites. We compare the obtained dispersion relation to the one obtained by the Bethe ansatz \cite{Yang:PRB1966a,Yang:PRB1966b,Yang:PRB1966c,Korepin:QISM1993}.
Within the Bethe ansatz it is possible to construct a single particle solution for the Hamiltonian $\HH$ of \Eqref{eq:H_LL}. However, the corresponding
fermionic creation operators are not given by a linear base transformation of the operators appearing in \Eqref{eq:H_LL}, instead the transformation is
higly non-linear and non-trivial \cite{Korepin:QISM1993}. In addition, the creation operator for the Bethe ansatz states are not sitting at 
equidistantly spaced momentum values.
Instead, the moment values, the so--called spectral parameter, have to be obtained by solving the Bethe ansatz equations. Finally the dispersion relation
is a cosine dispersion relation. For half filled systems one obtains for the Fermi velocity \cite{Yang:PRB1966a,Yang:PRB1966b,Yang:PRB1966c,Korepin:QISM1993}
\begin{align}
 \vF &= \frac{ \pi \sin(2\eta)t}{ \pi -2\eta}  \label{eq:vF_BA}\\
    U &= -2 \cos(2\eta) t \,,
\end{align}
where $\eta$ is a parametrization for the interaction $U$. We are therefore comparing the results of out iMF to a dispersion of
\begin{align}
 \varepsilon(p) &= \vF \, \cos(p) \,.
\end{align}
As displayed in \Figref{fig:LL_vF} we see a nice agreement between the approaches. It is only when approaching the transition to a phase separated state
for $U\rightarrow -2t$ that we find significant deviations. In result we see the major advantage compared to iDFT: we get access to a meaningful parametrization
of the kinetic energy. Within iDFT the dispersion relation would always correspond to the non-interacting one as the local densities are fixed
to precisely half filling on each site and the corresponding Kohn-Sham potentials are identical to zero. 
It also extends Bethe ansatz based DFT \cite{Capelle:PRB2006,Eckern:PRB2008}
beyond a local density approximation and providing improved hopping elements.

\subsection{Identifying excited states}
\begin{figure}[t]
\centerline{\includegraphics[width=0.95\columnwidth,type=pdf,ext=.pdf,read=.pdf]{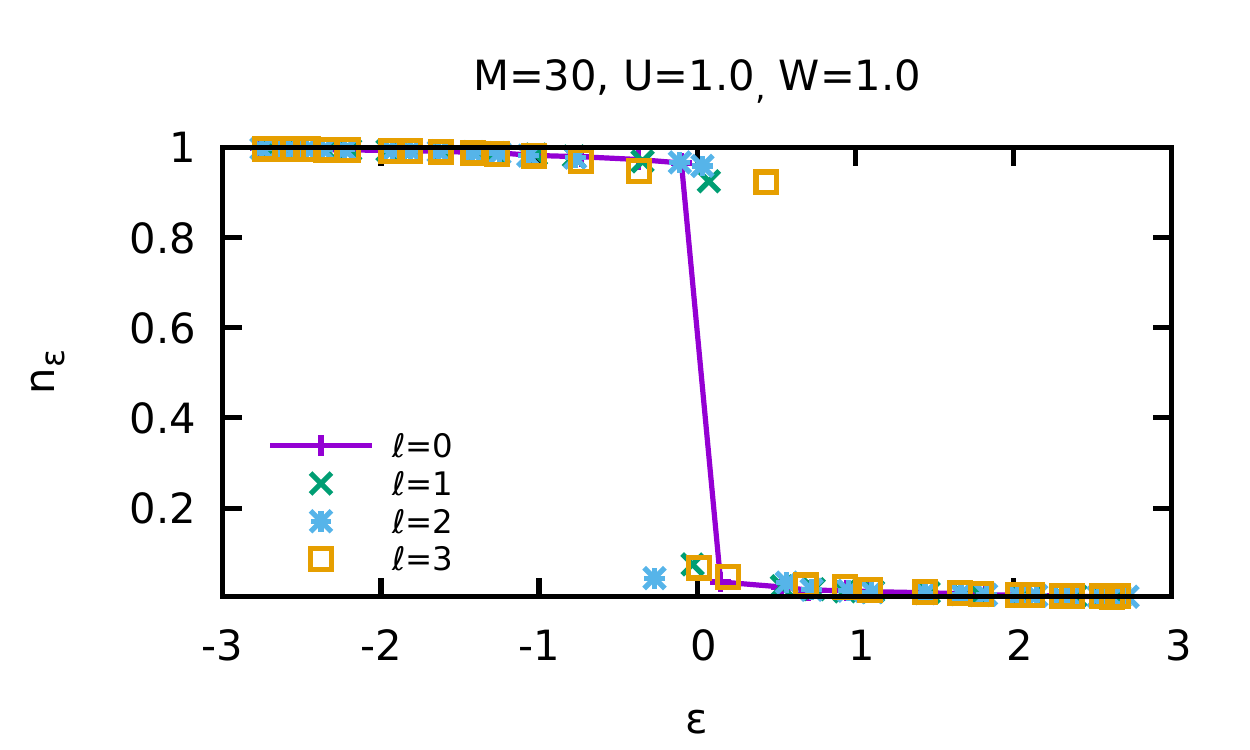}}
\caption{iMF occupation numbers --- Occupation of the iMF energy eigenstates for a system \Eqref{eq:LL_W} on $M=30$ sites with $t=1$, $U=1$, and $W=1.0$
 for the four states lowest in energy.}
\label{fig:ExLL}
\end{figure}

While DMRG is a powerful technique to study correlated quantum systems including measures for convergence,
there is actually one problem. The DMRG may converge to an excited state instead of true ground state.
In this situation the convergence measure of the DMRG will signal a nice convergence reflecting the fact
that on is indeed converged to the ground state. For details see Ref.~\cite{PS:Proceedings98,White:PRB05,PS:PRB2018}.
In this section we show that for fermionic systems the iMF approach can provide additional insight into this issue.
To this end we study a modle of a one--dimensional chain of fermions with nearest neighbor interaction and a box distributed on--site disorder

\begin{align}
\HH  =& -t \sum_x \hat{c}^+_{x-1} \hat{c}^{}_x  \,+\,\text{h.c.} \;+\;  \sum_x \epsilon^{}_x \hat{n}^{}_x \nonumber\\
     & + U \sum_x \left( \hat{n}_{x-1}-\frac{1}{2}\right) \left( \hat{n}^{}_x -\frac{1}{2}\right)  \label{eq:LL_W}
\end{align}

\begin{figure}[t]
\centerline{\includegraphics[width=0.95\columnwidth,type=pdf,ext=.pdf,read=.pdf]{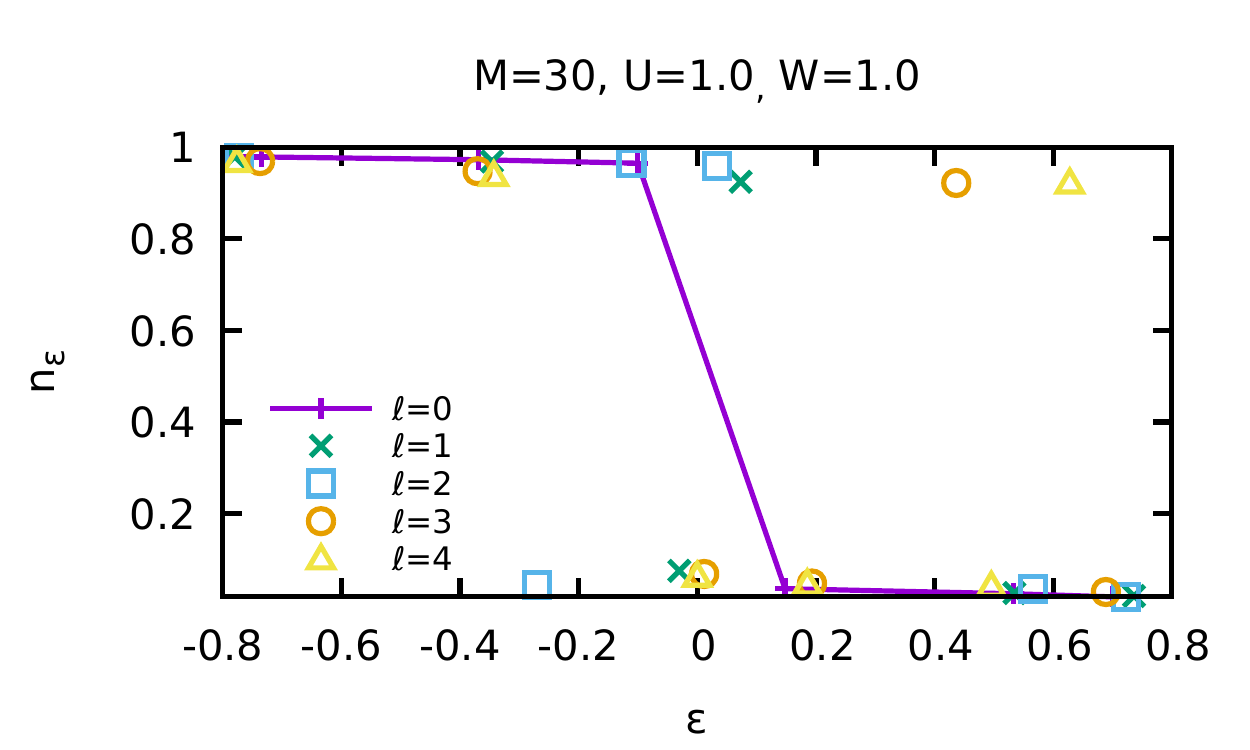}}
\caption{iMF occupation numbers --- A zoom closer to the Fermi point of the results provided in \Figref{fig:ExLL} for the five states lowest in energy.}
\label{fig:ExLL2}
\end{figure}

$\epsilon_x \in [ W/2, W/2$ of width $W$. Instead of extracting an iMF $\HiMF$ from the ground state obtained from the DMRG calculation,
we now extract an iMF Hamiltonian for each of the five states lowest in energy.
In \Figref{fig:ExLL} we show the occupation number, i.e. the diagonal element of $K$, vs.\ the energy eigenvalue of $\HiMF$ and in
\Figref{fig:ExLL2} we provide a zoom closer to the Fermi point. One can clearly see that only the ground state, $\ell=0$,  resembles
a monotonic distribution function. The excitation spectrum corresponding to excited many particle state $\ket{\Psi_\ell}$, $\ell > 0$,
show clear particle hole excitations. If our DMRG would have failed to provide the $\ell=0$ state then the non-monotonicity
of the distribution function vs.\ energy eigenvalue of the iMF Hamiltonian would provide a clear signal.
In addition, these results show that the energy eigenvalues of our iMF Hamiltonian do not depend strongly on the excitation level,
at least for these few low lying states which we take as a motivation to look into spectral functions in the following section.
%

\subsection{Side coupled resonant level model: spectral function} 

In this section we study a single impurity that is side coupled to a non-interacting one--dimensional lead,

\begin{align}\HH &= -t \sum_x \hat{c}^+_{x-1} \hat{c}^{}_x  \,+\, \text{h.c.} \;+\; \epsilon_{\mathrm d} \hat{n}_{\mathrm d}\\
    & - V \left(  \hat{c}^+_{x_0} \hat{d}  \,+\, \text{h.c.}  \right)  \;+\ U  (\hat{n}_{x_0} -1/2) ( \hat{n}_{d} - 1/2)  \,,
\end{align}
with $V$ the strength of the impurity coupling and $U$ a density--density interaction on the contact link.
In the non--interacting case, $U=0$, the system possess two bound states outside the conduction band we have
been investigated in context of radiation trapping in wave guiding structures %
\cite{ShenFan-2007,ShenFan-2007-1,LongoSchmitteckertBusch-2010,LongoSchmitteckertBusch-2011,Busch:PRA2016}. 

Resolving the existence of bound states close to the band edges is a numerically difficult task.
Within DMRG we applied a Chebyshev expansion as described in Ref.~\cite{Braun_PS:PRA2014}
for a system with $V=0.7$ and  $\epsilon_{\mathrm d}=0.5$ using 2000 Chebyshev moments.
For the numerical calculation we performed a symmetric and anti--symmetric combination of the lead sites,
$\hat{c}_{\pm,x} = \left( \hat{c}_{x} \pm \hat{c}_{-x}  \right)/\sqrt{2}$, $x>0$ and $x=0$ denoting the site that is coupled to the impurity.
Since the antisymmetric lead disconnects from the symmetric lead and the impurity, the model simplifies to a single impurity level
that couples to the end of a homogeneous tight binding chain with hopping $t=1$, except of the coupling between the first and the second chain sites, where
the hopping element is given by  $\sqrt{2} t$. In addition we applied damped boundary conditions (DBC) %
\cite{Vekic_White:PRB1993,BohrSchmitteckertWoelfle:EPL2006,Schmitteckert:JPCS2010,Braun_PS:PRA2014} 
where we apply a hopping of $t=1$ for the first six sites of a $M_{\mathrm L}=89$ sites chain, except the first hopping of the chain being $\sqrt{2}t$.
we then decrease the hopping element by factor of $\Lambda=0.98$ for the next 70 sites keeping the hopping element constant on the remaining bonds.  
In total the systems consists of $M=M_{\mathrm L}+1=90$ sites. The reason for choosing these DBCs is to ensure the necessary
energy resolution which at the same time makes these calculations pretty expensive as we kept up to 4500 states per DMRG block.

\begin{align}
\HH &= -t \sum_x \hat{c}^+_{x-1} \hat{c}^{}_x  \,+\, \text{h.c.} \;+\; \epsilon_{\mathrm d} \hat{n}_d\\
    & - V \left(  \hat{c}^+_{x_0} \hat{d}  \,+\, \text{h.c.}  \right)  \;+\; U  (\hat{n}_{x_0} -1/2) ( \hat{n}_{d} - 1/2) 
\end{align}
\begin{figure}
{\includegraphics[width=0.9\columnwidth,type=png,ext=.png,read=.png]{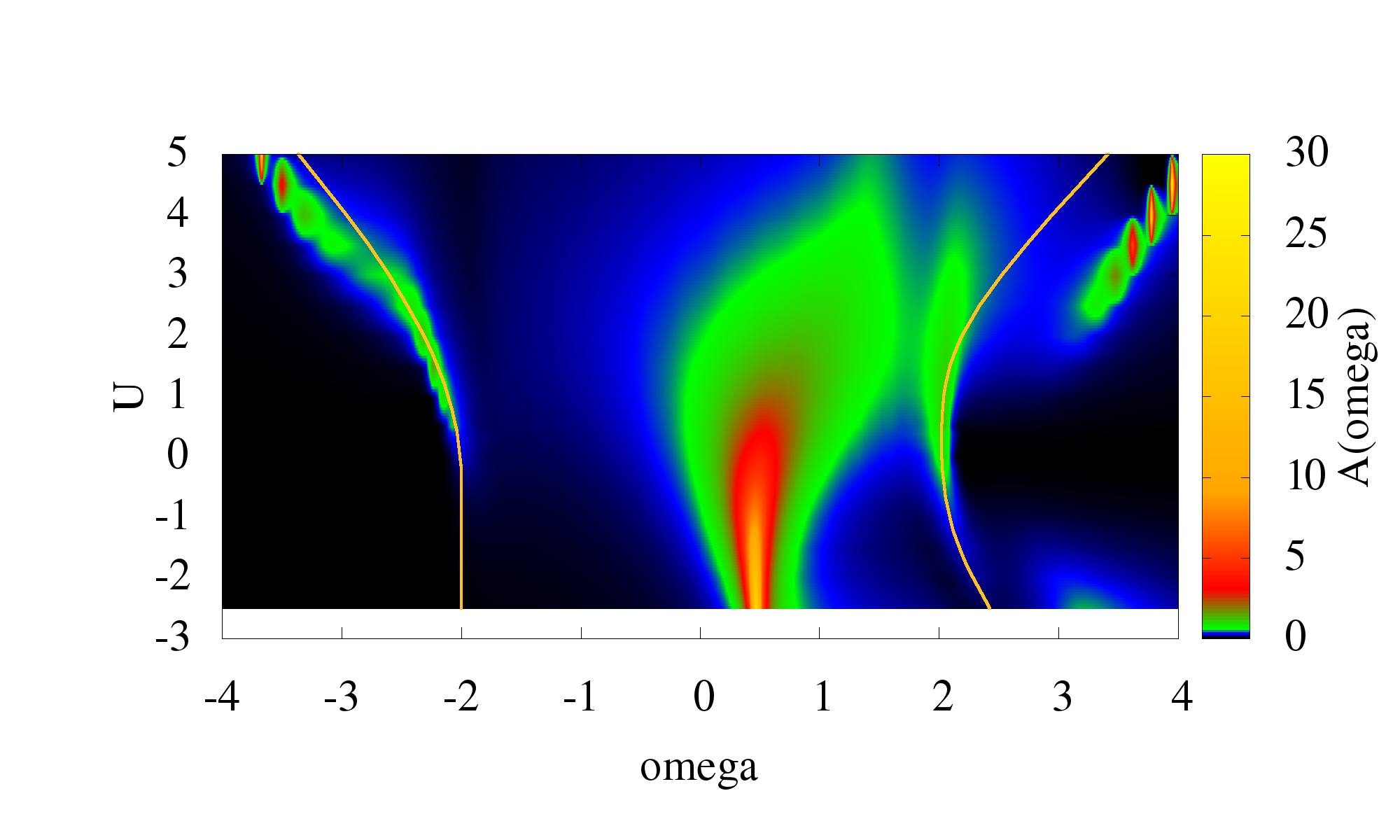}}
\caption{Spectral function of a side coupled impurity --- The spectral function $A(\omega$) as  a color coded plot vs. the frequency $\omega$ and the 
interaction $U$ on the contact link. The lines are obtained from the outermost eigenvalues of  the corresponding iHF provided thei energy is not within 
the support of the band $[-2, 2]$.}
\label{fig:SCRLM0}
\end{figure}

In \Figref{fig:SCRLM0} we display a color coded plot of the resulting impurity spectral function for a hybridization of $V=0.7$ and an impurity potential of
$\epsilon_{\mathrm d}=0.5$. On can clearly see the peak associated with the resonant level. In addition on sees the excitation corresponding to
the bound states outside the conduction band which disappears for attractive interaction only on the negative boundary of the lead.
Note that the impurity breaks the particle hole symmetry of the system.

Interestingly, if we extract the outermost energy levels of an iMF Hamiltonian corresponding to a 90 site system without employing DBC, leading to tremendously simplified numerics,
we find that we obtain a qualitative description of the bound states at the band edges. In the case of attractive interaction the lowest energy eigenvalue of $\HiMF$ is
above the lower band edge of $-2t$ in accordance with the nonexistence of a bound state in the spectral function in the regime.

\section{Kinetic bond MF}

Inspired by the generalized  current density function theory of \textcite{Tokatly:PRB2011} we can take 
our  iMF Hamiltonian as a starting point to search for a non--interacting Hamiltonian $\HH^0  = \sum_{x,y} \,H_{x,y}\, \hat{c}^+_x \hat{c}_y $
\begin{align}
  \bra{ \Psi } \hat{c}^+_x  \hat{c}^{}_x \ket{\Psi} &= \bra{ \Psi_0 } \hat{c}^+_x  \hat{c}^{}_x \ket{\Psi_0}   \label{eq:kbMF_n} \\
  t_{x,y} \bra{ \Psi } \hat{c}^+_x  \hat{c}^{}_y \ket{\Psi} &= H_{x,y} \bra{ \Psi_0 } \hat{c}^+_x  \hat{c}^{}_y \ket{\Psi_0}   \qquad x\ne y \label{eq:kbMF_T} \,,
\end{align}
where $\ket{\Psi}$ ($\ket{\Psi_0}$) is the many particle eigenstate of the interacting (non--interacting mean field) system,
$t_{x,y}$ the hopping elements of the interacting system and $H_{x,y}$ the matrix elements of the non-interacting system.
\Eqref{eq:kbMF_n} corresponds to the standard DFT condition, and  \Eqref{eq:kbMF_T} requires the kinetic energy of a bond of the interacting system
to equal the kinetic energy of a bond in the non-interacting system. Having solved the DFT equation  \Eqref{eq:kbMF_n} the kinetic part
of the MF system is then given
\begin{align}
    H_{x,y} &= t_{x,y} \frac{ \bra{ \Psi } \hat{c}^+_x  \hat{c}^{}_y \ket{\Psi}}{\bra{ \Psi_0 } \hat{c}^+_x  \hat{c}^{}_x \ket{\Psi_0}}   \qquad x\ne y \label{eq:kbMF_iT} \,.
\end{align}
We can now iterate the procedure of adapting local potentials to fulfill \Eqref{eq:kbMF_n}, obtaining the kinetic energy part from \Eqref{eq:kbMF_iT}.
Since we are adapting the kinetic energy of a bon we call it a kinetic bond mean field (kbMF).
\begin{figure}[t]
\centerline{\includegraphics[width=0.95\columnwidth,type=pdf,ext=.pdf,read=.pdf]{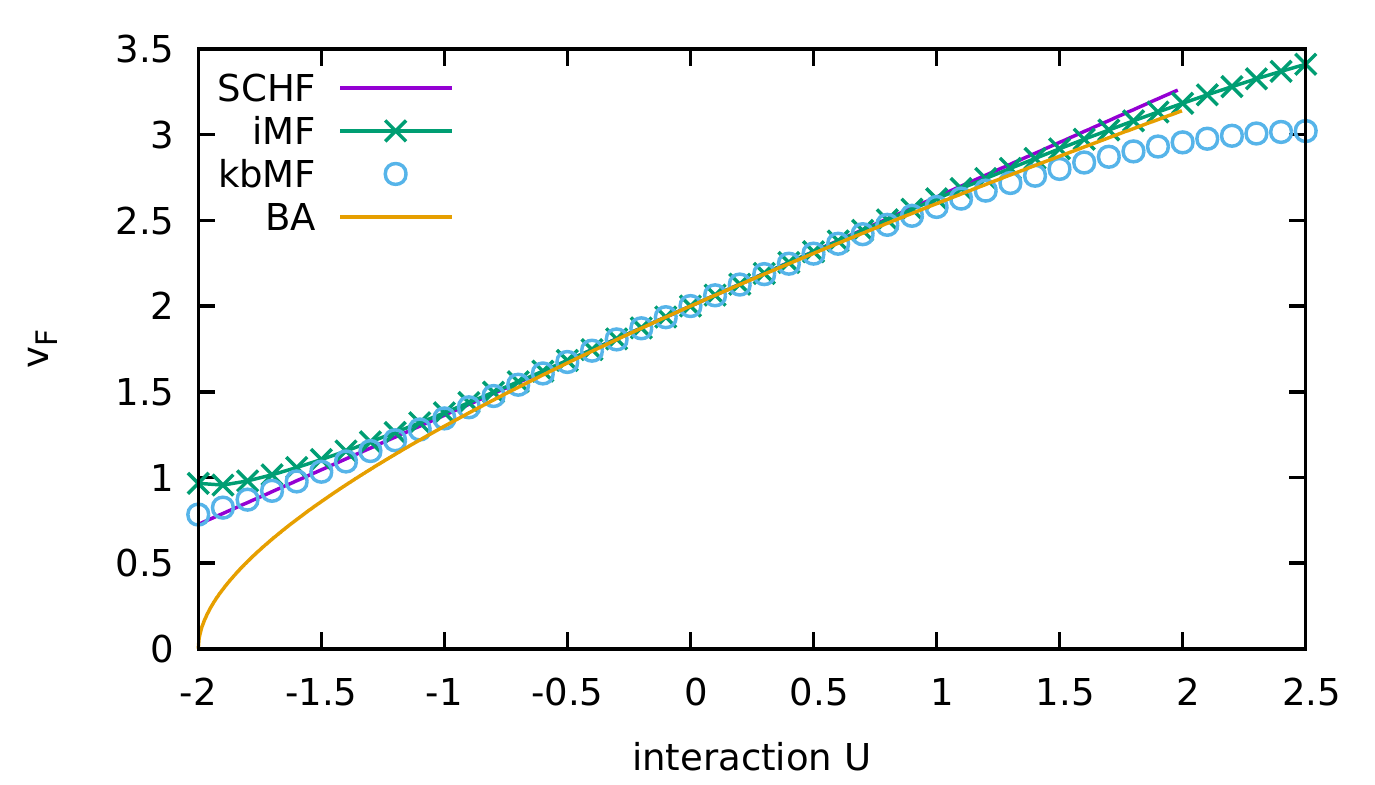}}
\caption{iMF Fermi velocities --- Fermi velocities obtained from SCHF (line), iMF (crosses), and kbMF (circles) in comparison
    of the Bethe ansatz result in the thermodynamic limit. The SCHF is performed on $M=30$ sites and the iMF calculation are performed on $M=30$
    where periodic boundary condition are employed.}
\label{fig:iMF_vF}
\end{figure}

In \Figref{fig:iMF_vF} we compare the Fermi velocity $\vF$ as obtained from the slope of the the single particle dispersion relation
in comparison the Bethe ansatz result \Eqref{eq:vF_BA}. In contrast to the Friedel oscillations in section \ref{sec:FO} the interaction
induced change of $\vF$ is already captured by SCHF for moderate interaction values. However, this result is misleading, as any small scatterer
will lead to an ordered state in SCHF, in contrast to our iMF descriptions.
Note that in contrast to the Bethe ansatz DFT \cite{Eckern:PRB2008} we obtain the $\vF$ already from the dispersion relation without resorting
to a time dependent approach. It should therefore provide a better starting point for a time dependent approach to obtain an improved value for $\vF$. 
The kbMF proved the best results in the regime of $U$ between $-1.5$ and $1.0$, while the iHF performs best for larger interaction values up to $U=2.0$.
These findings show that one should not blindly follow the recipes provided in this work. Instead one should check whether the mean field theory
can describe quantities of interest before interpreting the results. it also show the freedom in designing inverse mean field theory. We can pick a specific
property of the system and search for a non--interacting Hamiltonian reproducing this property.

\section{Summary}

In summary we have shown that one can extract meaningful single particle Hamiltonians from the many particle wave function
even for interacting one--dimensional Fermi systems where mean field theories are usually not justified.
Since we start from a many particle eigenstate one may ask whether there is actually any use of this approach as the problem appears to solved already.
First, as we pointed out before, it is hard to discuss physics via many particle wave functions.
A non--interacting picture may provide more insight.
Second, our approach provides a constructive route towards RDMFT beyond perturbative approaches.
Third, we hope that this approach can be used for the upscaling of numerics for correlated quantum systems.
An interesting feature of our calculations is that the resulting iMF Hamiltonian are pretty local in the sense that bonds
not appearing in the interacting Hamiltonian never get a significant contribution. Whether this is a general feature or just a coincidence
of our examples is an open question. 
And third, the iMF can be used to answer a tricky question arising in DMRG simulations, namely whether one has converged to a the ground or an excited state.
A question that can't be answered within current approaches.

Of course, one should expect that a non--interacting description can replace a full interacting treatment.
However, the extraction of a mean field theory that captures interaction effects  at least partially  may provide an improved starting point
for perturbative methods like functional renormalization group calculations \cite{Wetterich:PLB1993}.
We would also like to remark that \Eqref{eq:DOOFT_H} can be seen as the first order term of the resolvent for the retarded Greensfunction.
It should therefore be possible to extend the approach presented in this work to a Greens function based inverse mean field method
and a possible relation to such functional theories \cite{Potthoff:PRB2013,Schade:EPJST2017} is subject to further research.
Indeed, a time dependent extension of the approach presented is straightforward.

\section{Acknowledgement}
This work was supported by ERC-StG-Thomale-TOPOLECTRICS-336012.

\bibliographystyle{apsrev-nourl}
\bibliography{References}

\end{document}